\documentclass[conference]{IEEEtran}
\IEEEoverridecommandlockouts
\usepackage{cite}
\usepackage{amsmath,amsthm,amssymb,amsfonts}
\usepackage{algorithm}
\usepackage{algorithmic}  
\usepackage{graphicx}
\usepackage{textcomp}
\usepackage{xcolor}
\usepackage{bm}
\usepackage{subfigure}
\usepackage{setspace}
\def\BibTeX{{\rm B\kern-.05em{\sc i\kern-.025em b}\kern-.08em
    T\kern-.1667em\lower.7ex\hbox{E}\kern-.125emX}}
\begin{document}
\title{\vspace{-10mm}Noisy Sensor Scheduling in Wireless Networked Control Systems: Freshness or Precision 
\vspace{-4mm}}

\author{He~Ma~and~Shidong~Zhou,~\IEEEmembership{Member,~IEEE} \vspace{-8mm}
\thanks{This work is supported by Key Research and Development Program of China under Grant 2018YFB1801102, Tsinghua University-China Mobile Communications Group Co.,Ltd. Joint Institute, National Natural Science Foundation of China under Grant 61631013, Foundation for Innovative Research Groups of the National Natural Science Foundation of China under Grant 61621091, Tsinghua University-China Mobile Communications Group Co.,Ltd. Joint Institute, Huawei Company Cooperation Project under Contract No. TC20210519013.
}
}
\maketitle

\begin{abstract}
In linear wireless networked control systems whose control is based on the system state's noisy and delayed observations, an accurate functional relationship is derived between the estimation error and the observations' freshness and precision. The proposed functional relationship is then applied to formulate and solve the problem of scheduling among different wireless links from multiple noisy sensors, where a sliding window algorithm is further proposed. The algorithm's simulation results show significant performance gain over existing policies, even in scenarios that require high freshness or precision of observations.
\end{abstract}

\vspace{-6mm}
\section{Introduction}
\vspace{-2mm}
Networked Control Systems (NCSs) are vital in industrial automation and the Internet of Things with applications ranging from the chemical industry and autonomous driving to smart cities and smart grids\cite{gopalakrishnan2011incorporating,zhang2015survey}.
 
With increasing demands of mobility and flexibilities, more and more connections in NCSs need to be replaced by wireless links, forming Wireless Networked Control Systems (WNCSs) \cite{park2017wireless}. 
However, data transmission through wireless links can not be as reliable and timely as wired links since only part of the links can transmit successfully due to the limited resource and the unexpected time-varying fading channel\cite{ shafi20175g}.
Thus radio resource scheduling is needed to ensure control performance.

Proper scheduling design calls for the evaluation of nonideal communication's impact on the performance.
\cite{kim2003maximum} has considered maximum delay as a criterion and researched the relationship between delay and control stability. 
Moreover, the maximum allowable transmission interval has been proposed to explain how the sampling rate affects stability in \cite{carnevale2007lyapunov}. 
 
Stability is only one primary target of a control system. More efforts are needed to approach higher control accuracy.
However, it is hard to establish a functional relation between communication performance and control accuracy.
In WNCSs, state estimation is important as the control decision might only be made based on imperfect observations of the system state. Therefore, the state estimation error has been taken as an indirect optimization target in \cite{champati2019performance}, \cite{ayan2019age}, and \cite{kiekenap2020optimum}, where the relationship between the estimation error and the Age of information\cite{sun2017remote} at the controller has been studied. 
{However, in \cite{champati2019performance,ayan2019age,R2,R3,R4,R1}, only the impact of delay and packet dropouts is discussed. Meanwhile, noisy observations with no time delay in WNCSs are studied in \cite{R5}. Methods used in these works are not enough for noisy and delayed observations as decisions need to be made among fresh observations with low precision and stale observations with high precision.}
 
{Mamduhi and Champati\cite{mamduhi2020freshness} considered noisy and delayed observations in WNCSs. However, only some heuristic scheduling policies are proposed without the evaluation function of noisy and delayed observations, which can further help optimize the control performance. The functional impact of both the delay process and noise variance process on the state estimation error in WNCSs is important but complex, and therefore mainly considered in the scalar state case at present. Singh and Kamath\cite{singh2019optimal} provided the pioneering work on the functional impact in scalar WNCSs with multiple sensors, which is very helpful for scheduling design in WNCSs. }

However, the relationship in \cite{singh2019optimal} between estimation error and delay is underestimated since the correlation between the estimation error and the historical process noise is ignored. A more accurate evaluation of the estimation error is needed. 

{In this letter, we will still focus on scalar WNCSs with multiple sensors and find a more direct and accurate           interface between control performance and communication performance. The interface will take delay and observation noise into account and reflect control performance. }The main contributions of this letter include:
\vspace{-1mm}
\begin{itemize}
\item An accurate functional relationship has been established between the communication performance process (delay and variance of updated observation noise) and the time-average estimation squared error as well as the control cost for multi-sensor scalar WNCSs with noisy observations. The main difference between the proposed function and the function in \cite{singh2019optimal} is discussed. 
\item A sliding window algorithm is proposed to schedule multiple sensors by using the proposed function for a compromise between complexity and performance. 
\end{itemize}
\vspace{-2mm}
\section{System Model}

\begin{figure}[htbp]
\centerline{\includegraphics[width=3in]{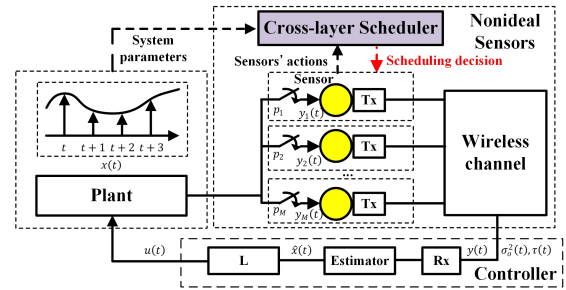}}
\vspace{-3mm}
\caption{System Model}
\label{fig}
\vspace{-3mm}
\end{figure}

\vspace{-2mm}
In this paper, we focus on the WNCSs 
shown in Fig.\ref{fig}.
\vspace{-2mm}

It is assumed that time has been discretized into time slots $t$ with the same length. The system consists of a plant, multiple nonideal sensors with different precisions and a controller. In every time slot $t$, the controller receives a delayed noisy observation of the system state from the wireless channel. 
\vspace{-2mm}
\subsection{System dynamic}
\vspace{-1mm}
The system dynamic is considered as a scalar Gauss Markov process with control input which evolves as: 
\vspace{-1mm}
 \begin{equation}\label{control model}
\begin{aligned}
x(t+1) &=ax(t)+bu(t)+w(t)\\
\end{aligned}
\vspace{-1mm}
\end{equation}
where $x(t)\in\mathbb{R}$, $u(t)\in\mathbb{R}$ represent the system state and control input, respectively. Constant scalars $a\in\mathbb{R}$ and $b\in\mathbb{R}$ describe the system and input coefficient, respectively. The process noise $w(t)$ is a zero-mean Gaussian noise with variance ${\sigma}_p^2$, which is i.i.d. across time. 
\vspace{-2mm}
\subsection{Controller and Estimator}
\vspace{-1mm}
The controller uses the real-time estimates $\hat{x}(t)$ to produce a control input $u(t)$. 
Without loss of generality, we adopt a linear control strategy as follows:
\vspace{-1mm}
 \begin{equation}\label{control strategy}
\begin{aligned}
u(t) &= L\hat{x}(t)          
\end{aligned}
\vspace{-1mm}
\end{equation}
where $L$ is the feedback control gain.

Delayed Kalman filter is usually applied for state estimation with delayed and imperfect observations\cite{kumar2015stochastic}. Assuming that at time slot $t$ the received observation $y(t) = x(t-\tau(t))+v(t-\tau(t))$ has the time delay $\tau(t)$ and the  zero-mean Gaussian noise $v(t-\tau(t))$ with variance $\sigma^2_o(t)$, the estimation therefore can be updated as follows: 
\vspace{-1mm}
 \begin{equation}\label{estimation model}
\begin{aligned}
 \hat{x}(t) &=\hat{x}^{-}(t)+k(t)(a^{{\tau}(t)}y(t)-\hat{x}^{*}(t))   
\end{aligned}
\end{equation}
where $k(t)$ is the estimation gain, $\hat{x}^{-}(t)=a\hat{x}(t-1)+bu(t-1)$ is the one step prediction of the state and $\hat{x}^{*}(t)$ is the ${\tau}(t)$ steps state compensation. When ${\tau}(t)=0$, $\hat{x}^{*}(t)=\hat{x}^{-}(t)$ and when ${\tau}(t)>0$, $\hat{x}^{*}(t)=a\hat{x}(t-1)-\sum^{{\tau}(t)-1}_{l=1}a^lbu(t-1-l)$. 

\vspace{-2mm}
\subsection{Nonideal sensors}
\vspace{-1mm}
{ Due to unpredictable error and interruptions, a set of sensors $\mathcal{M}=\{1,...,M\}$ are assumed to take probabilistic noisy observations $y_m(t)$ of the system state $x(t)$  with probability $p_m$ to ensure the control performance.}
\vspace{-1mm}
\begin{equation}\label{sensor model 1}
\begin{aligned}
y_m(t)=   x(G_m(t))+v_{m}(G_m(t))
\end{aligned}
\vspace{-1mm}
\end{equation}
where $G_{m}(t)$ is the generation time of the latest observation produced by the sensor $m\in\mathcal{M}$ at slot $t$ and the observation noise $v_{m}(t)$ is a zero-mean Gaussian noise with variance $\sigma^2_{o,m}$, which is i.i.d across time and different sensors. The observe probability is assumed to be time-invariant for simplification.
\vspace{-2mm}
{\subsection{Communication model}}
\vspace{-1mm}
{Sensors send state observations, in the form of packets, to the controller over a shared wireless channel.  The wireless channel is assumed to be error-free but bandwidth-limited, i.e. one packet per time slot.  The transmission delay from the scheduled sensor's side to the controller is denoted as $t_d$, which is assumed to be fixed for simplification.
The transmission activity is controlled by a scheduler that makes the transmission scheduling decisions $d(t)\in\mathcal{M}$ from all sensors with information of the system parameters $a,b,\sigma^2_p,\sigma_{o,m}^2$ and sensors' action (sense or not) at each slot. }Note that the scheduler does not need instantaneous information of the system state. Instead, only the traffic information and the static system parameters are needed.

Then, the age of the controller's received observation $y(t)$ is defined as: 
\vspace{-1mm}
 \begin{equation}\label{AoI model}
\begin{aligned}
             \tau(t) = t - G_{d(t-t_d)}(t-t_d)
\end{aligned}
\vspace{-1mm}
\end{equation}
{where $d(t-t_d)$ is the sensor number corresponding to the state observation received by the controller in the current time slot $t$.}
Therefore the noise variance of the received observation at slot $t$ in the controller has $\sigma^2_o(t) = \sigma^2_{o,d(t-t_d)}$. {The age of observations mainly comes from irregular observing, untimely scheduling and the transmission delay.}
\vspace{-1mm}
\section{State Estimation Error Analysis}
\vspace{-1mm}
As the controller's estimation error determines control performance, it is important to derive the functional relationship between estimation error and the processes of delay and precision of the received state observation at the controller.

This section provides a general theorem on the functional relationship and shows the difference from that in \cite{singh2019optimal}.
{Note that the analysis in this section does not rely on the assumption about the sensor model and communication model.}
To evaluate estimation performance, the time-average squared error of state estimation \cite{aastrom2012introduction} is defined as follows:
\vspace{-1mm}
 \begin{equation}\label{MSE_SE}
\begin{aligned}
             J_E(T) =\frac{\mathbb{E}[\sum^{T-1}_{t=0}e^2(t)]}{T}= \frac{\sum^{T-1}_{t=0}\mathbb{E}[(x(t)-\hat{x}(t))^2]}{T}
\end{aligned}
\vspace{-1mm}
\end{equation}
$\boldsymbol{\tau}_t=\{\tau(0),\tau(1),...,\tau(t)\}$ and $\boldsymbol\sigma^2_{t}=\{\sigma^2_{o}(0),\sigma^2_{o}(1),...,\sigma^2_{o}(t)\}$ are defined to represent the age process and observation noise variance process of received observations from time slot $0$ to time slot $t$, respectively. The following functional expression is derived to evaluate the impact of these processes on the time-average squared error of state estimation. 
\vspace{-1mm}
\newtheorem{theorem}{Theorem}
\begin{theorem} \label{lemmaPerform}
In a discrete-time control system as described in \eqref{control model}, where the controller makes linear control \eqref{control strategy} based on the state estimation, which is estimated by the Kalman filter \eqref{estimation model} with the received delayed and noisy observation, the time-average squared estimation error \eqref{MSE_SE} has the form as follows:
\vspace{-2mm}
 \begin{equation}\label{control metric}
\begin{aligned}
J_E(T) &=J_E(\boldsymbol{\tau}_T,\boldsymbol{\sigma}^2_T)\\
& = \frac{1}{T}\sum_{t=0}^{T-1}\sum_{q=t}^{T}\prod_{n=1}^{q-t}[a^2{(1-k(t+n))}^2]f(t,\sigma^2_{o}(t),\boldsymbol{\tau}_t)
\end{aligned}
\end{equation}
where $f(t,\sigma^2_{o}(t),\boldsymbol{\tau}_t)$ is the core function affected by the nonideal observation and transmission as: 
 \begin{equation}\label{reward function fuza}
\begin{aligned}
f(t,\sigma^2_{o}(t),\boldsymbol{\tau}_t) =&C_p(t,\tau(t))+C_s(t,\tau(t),\sigma^2_{o}(t))+C_e(t,\boldsymbol{\tau}_t)\\
C_p(t,\tau(t))=&{\sigma}_p^2+\sum^{\tau(t)-1}_{l=1}a^{2l}k^2(t){\sigma}_p^2\\
C_s(t,\tau(t),\sigma^2_{o}(t))&=k^2(t)a^{2\tau(t)}\sigma^2_{o}(t)\\
C_e(t,\boldsymbol{\tau}_t)&=2\sum^{\tau(t)}_{l=2}a^{l-1}k(t)\prod_{j=0}^{l-2}{\left[a(1-k(t-j))\right]}{\sigma}_p^2+\\
&2\sum^{\tau(t)-2}_{n=1}k(t-n)k(t)\prod_{j=0}^{n-1}{[a(1-k(t-j))]}\\ &\sum^{\min\{\tau(t)-n,\tau(t-n)\}-1}_{l=1}a^{2l+n}{\sigma}_p^2
\end{aligned}
\vspace{-4mm}
\end{equation}
\begin{proof}
From Equations \eqref{control model} and \eqref{estimation model},  the estimation error $e(t)$ evolves as follows:
\begin{equation}\label{Linear Quadratic cost 3}
\begin{aligned}
e(t)  &= a(1-k(t))e(t-1)-k(t)a^{\tau(t)}v(t-\tau(t))\\
&+[w(t-1)+k(t)\sum^{\tau(t)-1}_{l=1}a^lw(t-l-1)]
\end{aligned}
\end{equation}
Therefore, the estimation variance at time slot $t$ is given by:
\begin{equation}\label{Linear Quadratic cost 4}
\begin{aligned}
\mathbb{E}\{e^2(t)\}  &= a^2(1-k(t))^2\mathbb{E}\{e^2(t-1)\}+\\
&k^2(t)a^{2\tau(t)}\sigma^2_{o}(t)+({\sigma}_p^2+\sum^{\tau(t)-1}_{l=1}a^{2l}k^2(t){\sigma}_p^2)\\
+2&\mathbb{E}\left[a(1-k(t))e(t-1)k(t)\sum^{\tau(t)-1}_{l=1}a^lw(t-l-1)\right]
\end{aligned}
\end{equation}
where the last terms indicates that the estimation error $e(t-1)$ is correlated with the historical process noise. Knowing that the process and observation noise are independent across time, we have from Equation \eqref{Linear Quadratic cost 3} that:
\begin{equation}\label{Linear Quadratic cost 5}
\begin{aligned}
&\mathbb{E}\left[e(t-1)\sum^{\tau(t)-1}_{l=1}a^lw(t-l-1)\right]  \\
&= \mathbb{E}\{[a(1-k(t-1))e(t-2)+w(t-2)+k(t-1)\\
&\sum^{\tau(t-1)-1}_{l=1}a^lw(t-l-2)]a[w(t-2)+\sum^{\tau(t)-2}_{l=1}a^{l}w(t-l-2)]\}  \\
&= \mathbb{E}[a(1-k(t-1))e(t-2)\sum^{\tau(t)-1}_{l=1}a^lw(t-l-1)]+\\
&a\sigma^2_p+k(t-1)\sum^{\min\{\tau(t-1),\tau(t)-1\}-1}_{l=1}a^{2l+1}\sigma^2_p\\
\end{aligned}
\end{equation} 
It can be observed that $e(t-1)$ is correlated with $w(t-2)$ and $\sum^{\tau(t)-2}_{l=1}a^{l}w(t-l-2)$. Therefore if the correlation is ignored, the estimation error will be underestimated.

By iteratively executing the expansion in \eqref{Linear Quadratic cost 5}, the form $C_e(t,\boldsymbol{\tau}_t)$ can be derived and we can furthermore express $\mathbb{E}\{e^2(t)\}$ as the following form:
\begin{equation}\label{Linear Quadratic cost 6}
\begin{aligned}
\mathbb{E}\{e^2(t)\}  =& a^2(1-k(t))^2\mathbb{E}\{e^2(t-1)\}+f(t,\sigma^2_{o}(t),\boldsymbol{\tau}_t)
\end{aligned}
\end{equation}
where $f(t,\sigma^2_{o}(t),\boldsymbol{\tau}_t)$ is shown in Equation \eqref{reward function fuza}.

Then by iteratively substituting Equation \eqref{Linear Quadratic cost 6} into \eqref{control metric} and calculating the sum of the series, the proof is complete.
\end{proof}
\vspace{-2mm}
\end{theorem}
\vspace{-4mm}
The meanings of each term are as follows. 

$C_p(t,\tau(t))$ represents the effect brought by the process noise. $C_p(t,\tau(t))$ increases as the variance of the process noise $\sigma_p^2$ and the sensor's age $\tau$ increase. 

$C_s(t,\tau(t),\sigma^2_{o}(t))$ represents the effect brought by the observation noise and is mainly determined by the variance of the observation noise $\sigma^2_{o}(t)$ and age $\tau$. 

$C_e(t,\boldsymbol{\tau}_t)$ represents the correlation between the historical process noise and the estimation error of the last slot $e(t-1)$. 

$f(t,\sigma^2_{o}(t),\boldsymbol{\tau}_t)$ is linear with the variance of process noise $\sigma_p^2$ and observation noise $\sigma^2_{o}(t)$. The age contributes to the performance function in the form of an exponential power with base $a$. The observation therefore needs both freshness and precision to obtain better performance.
The theorem provides a numerical evaluation criterion of delayed and noisy observations, which can guide communication design in WNCSs.

The following corollary can be derived directly under a special case where $k(t) = k$ and $|a(1-k)|<1$ for the convergence of the estimation process.
\vspace{-2mm}
\newtheorem{corollary}{Corollary}
\begin{corollary} 
In WNCSs as described in Theorem \ref{lemmaPerform}, if the estimation gain $k$ is a constant and follows $|a(1-k)|<1$, then the infinite horizon time-average squared estimation error has the form as follows:
\vspace{-1mm}
 \begin{equation}\label{control metric3}
\begin{aligned}
 \lim_{T \to \infty}J_E(T) = \lim_{T \to \infty}\frac{\sum_{t=0}^{T-1}\mathbb{E}\{\frac{1}{1-{(a(1-k))}^2}f(t,\sigma^2_{o}(t),\boldsymbol{\tau}_t)\}}{T}\\
\end{aligned}
\end{equation}
where $f(\sigma^2_{o}(t),t,\boldsymbol{\tau}_t)$ is the core function affected by the nonideal observation and transmission as:
 \begin{equation}\label{reward function fuza1}
\begin{aligned}
f(t,\sigma^2_{o}(t),\boldsymbol{\tau}_t) &=C_p(\tau(t))+C_s(\tau(t),\sigma^2_{o}(t))+C_e(t,\boldsymbol{\tau}_t)\\
C_p(\tau(t))&=\frac{k^2\sigma_p^2(a^{2\tau(t)}-a^2)}{a^2-1}+\sigma^2_p\\
C_s(\tau(t),\sigma^2_{o}(t)&)=k^2{\sigma^2_{o}(t)}a^{2\tau(t)}\\
C_e(t,\boldsymbol{\tau}_t)&=\frac{2[(a^2(1-k))^{\tau(t)}-a^2(1-k)]k\sigma_p^2}{a^2(1-k)-1}+\\
2\sum_{n=1}^{\tau(t)-2}&\frac{k^2\sigma_p^2(a^2(1-k))^n}{a^2-1}(a^{2\min{\{\tau(t)-n,\tau(t-n)\}}}-a^2)
\end{aligned}
\vspace{-4mm}
\end{equation}
\label{cor-1}
\end{corollary}
\vspace{-2mm}
Note that the functional relationship in \cite{singh2019optimal} only contains $C_p(\tau(t))$ and $C_s(\tau(t),\sigma^2_{o}(t))$ in Corollary 1. 
\vspace{-1mm}
\section{Scheduling Problem Formulation}
\vspace{-1mm}
To minimize Linear Quadratic Gaussian (LQG) cost \cite{aastrom2012introduction}, the scheduling problem can be formulated as:
\begin{equation}\label{P1}
\begin{aligned}
&\min _{\pi}\,\,J(\pi):=\mathbb{E}_{\pi}\left(\lim_{T\to +\infty}\frac{1}{T}\mathbb{E}\left[\sum^{T}_{t=1}x^2(t)Q+u^2(t)R\right] \right) \\
&s.t.\quad d(t)\in\mathcal{M}, \forall t>0
\end{aligned}
\vspace{-1mm}
\end{equation}
where $Q\geq0$ and $R>0$ represent the state and control input weights respectively. The expectation is taken with respect to the randomness of the process and observation noises as well as the generic scheduling policy $\pi$.

According to theorem 1 in \cite{mamduhi2020freshness}, the optimal
control gain $L$ can be designed separately from the scheduling policy $\pi$ and is a certainty equivalent strategy \cite{aastrom2012introduction} as follows:
\vspace{-2mm}
 \begin{equation}\label{control1}
\begin{aligned}
L = -(Q+b^2P)^{-1}bPa          
\end{aligned}
\vspace{-2mm}
\end{equation}
where $P$ is the solution to \textit{Discrete Algebra Recatti equation} (DARE) \cite{aastrom2012introduction}.
Using the certainty equivalence property of \eqref{control1} in \cite{aastrom2012introduction}, we can rewrite the LQG control cost $J$ as follows:
\vspace{-2mm}
\begin{equation}\label{Linear Quadratic cost 1}
\begin{aligned}
J  &= c_0+\lim_{T \to \infty}\frac{1}{T}\sum_{t=0}^{T-1}\left[L^2(Q+b^2P)\right]\mathbb{E}\left[e^2(t)\right]\\
&=c_0+L^2(Q+b^2P)J_E
\end{aligned}
\vspace{-1mm}
\end{equation}
where $c_0 =  \mathbb{E}\{x^2(0)P\}+\sigma^2_pP$ is the initial condition by taking expectation on the randomness of the process noise.
Since $c_0$ and ${L^2(Q+b^2P)}$ in \eqref{Linear Quadratic cost 1} are irrelevant with the scheduling policy $\pi$, the original problem \eqref{P1} can be reduced to the following form according to Theorem \ref{lemmaPerform}:

\vspace{-3mm}
\begin{equation}\label{P2}
\begin{aligned}
&\min _{\pi} J_E(\pi)=\lim_{T \to \infty}\frac{1}{T}\sum_{t=0}^{T-1}\mathbb{E}\{\sum_{q=t}^{T}\prod_{n=1}^{q-t}[a^2{(1-k(t+n))}^2]\\&\quad\quad\quad\quad\quad\quad f(t,\sigma^2_{o,d(t)},\boldsymbol{\tau}_t)\}\\
&s.t.\quad d(t)\in\mathcal{M}, \forall t>0\\
\end{aligned}
\vspace{-2mm}
\end{equation}
\vspace{-4mm}
\section{Near-optimal policy design}\label{sec:policy}
\vspace{-1mm}
{
The scheduling problem, though as simplified as \eqref{P2}, is still difficult to be solved as a Markov Decision Process directly similar to \cite{R5}, since the function $f$ is a non-linear function of age, and there is coupling among the state estimation and process noise  from different time slots. Thus we try to provide an asymptotically optimal scheduling algorithm instead.}

Given the age process $\boldsymbol{\tau}_t$ of received observations, the age of all sensors $\boldsymbol{\Delta}=\{t-G_m(t)+t_d\}^M_{m=1}$ and estimation gain $k(t)$ at slot $t$, the following  $N$ step myopic problem is used to approximate the optimal policy by substituting $N$ into $T$:
\begin{equation}\label{P3}
\begin{aligned}
&\min _{\pi} J(\pi,t):=\frac{1}{N}\sum_{n=0}^{N-1}\mathbb{E}\{\sum_{q=n}^{N}\prod_{l=1}^{q-n}[a^2{(1-k(t+n+l))}^2]\\&\quad\quad\quad\quad f(t+n,\sigma^2_{o,d(t)},\boldsymbol{\tau}_{t+n})\} \\
&s.t.\quad d(t)\in\mathcal{M}, \forall t>0\\
\end{aligned}
\end{equation}
Instead of directly tackling the original problem \eqref{P2}, we obtain an approximate solution by solving the dynamic programming \eqref{P3} with a sliding window at each time slot. When $N$ approaches infinity, problem $\eqref{P3}$ is equal to problem $\eqref{P2}$, meaning that the solution is asymptotically optimal. While when $N=1$, the solution to problem $\eqref{P3}$ is the greedy policy. 

Consider the age of all sensors $\boldsymbol{\Delta}$ as the state of the dynamic programming problem \eqref{P3}. The main idea of the solution uses the Bellman equation \eqref{Bellman} to make backward inductions from time slot $t+N-1$ to time slot $t$.
\begin{equation}\label{Bellman}
\begin{aligned}
V&(n,\boldsymbol{\Delta},\boldsymbol{\tau}_{t+n})=\\
&\min_{d(n)}\sum_{\boldsymbol{\Delta}^{\prime}}Pr(\boldsymbol{\Delta}^{\prime}|\boldsymbol{\Delta},d(n))\{V(n+1,\boldsymbol{\Delta}^{\prime},\boldsymbol{\tau}_{t+n+1})\\&+\sum_{q=n}^{N}\prod_{l=1}^{q-n}[a^2{(1-k(t+n+l))}^2]f(t,\sigma^2_{o,d(n)},\boldsymbol{\tau}_{t+n})\}\end{aligned}
\vspace{-1mm}
\end{equation}
where $V(n,\boldsymbol{\Delta},\boldsymbol{\tau}_{t+n})$ is the value function of state $\boldsymbol{\Delta}$ with age process $\boldsymbol{\tau}_{t+n}$ and $Pr(\boldsymbol{\Delta}^{\prime}|\boldsymbol{\Delta},d(n))$ is the trasition probabiltiy from state $\boldsymbol{\Delta}$ to state $\boldsymbol{\Delta}^{\prime}$ under the action $d(n)$.  

We present Algorithm \ref{ALO1} to solve problem \eqref{P3}. 
\begin{algorithm}
 \caption{A Sliding Window Iteration Scheme}
 \label{ALO1}
 \begin{algorithmic}[1]
 \renewcommand{\algorithmicrequire}{\textbf{Input:}}
 \renewcommand{\algorithmicensure}{\textbf{Output:}}
 \REQUIRE The current age for all sensors and age process of received observations, $\boldsymbol{\Delta}$ and $\bm{\tau}_t$; the set of system parameters, $\mathbb{I}_s=\{a,\{k(t)\},\sigma^2_p,\sigma^2_{o,1},\ldots,\sigma^2_{o,M}\}$; 
 \ENSURE  the chosen sensor in the current time slot $t$, $d(t)$;  
  \FOR {$t \geq 0$}
  \STATE solve the $N$ step value iteration by \eqref{Bellman},
    \STATE choose the sensor which gets the minimal value function $V(t,\boldsymbol{\Delta},\boldsymbol{\tau}_{t})$ as $d(t)$ and \textbf{return} $d(t)$
   \ENDFOR
 \end{algorithmic} 
 \vspace{-1mm}
 \end{algorithm}
 {Rewards summed over multiple time steps are used to overcome the problem of shortsighted scheduling.  The Algorithm \ref{ALO1} is a trade-off between the LQG optimality and the complexity. } 
\vspace{-2mm}

\section{Numerical results}
\vspace{-1mm}
In this section, we evaluate the function in theorem \ref{lemmaPerform} and the performance of Algorithm \ref{ALO1} through simulations. 
\vspace{-2mm}
\subsection{Estimation Error Evaluation Accuracy}
\begin{figure*}[hbtp]
\centering
\subfigure[]{
\begin{minipage}[t]{0.31\linewidth}\label{fg:sub4}
\centering
\includegraphics[width=2.5in]{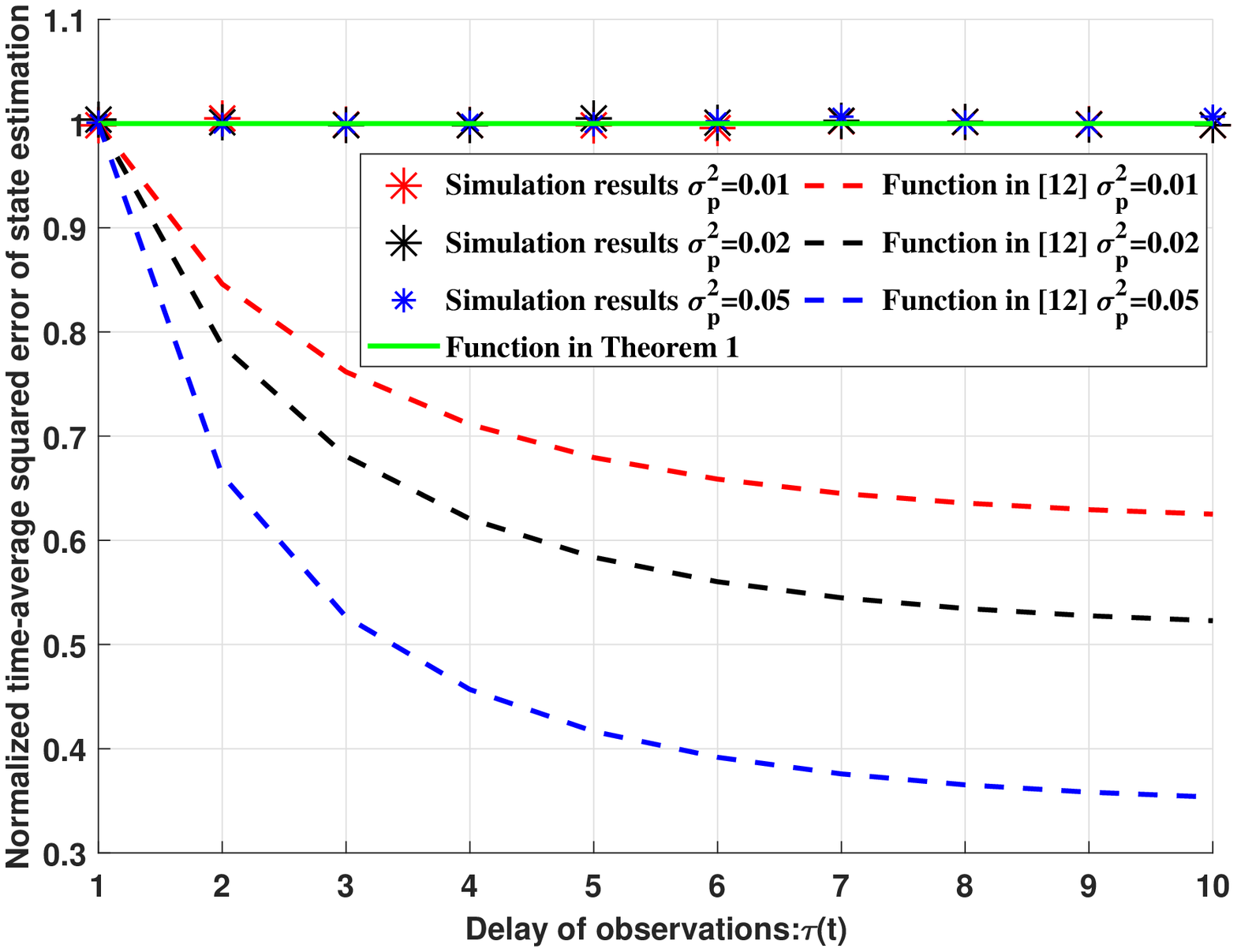}
\end{minipage}%
}
\subfigure[]{
\begin{minipage}[t]{0.31\linewidth}\label{fg:sub1}
\centering
\includegraphics[width=2.5in]{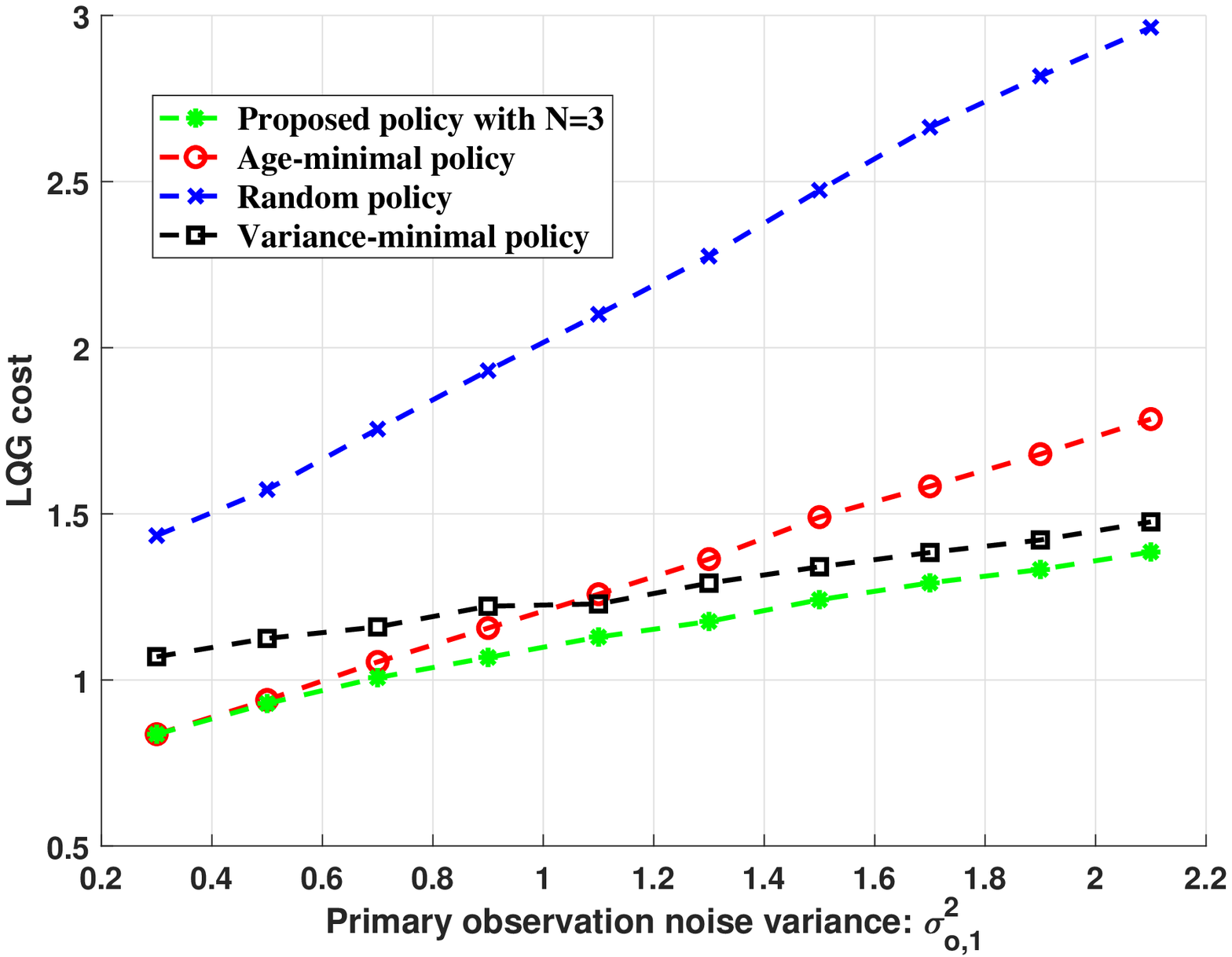}
\end{minipage}%
}%
\subfigure[]{
\begin{minipage}[t]{0.31\linewidth}\label{fg:sub2}
\centering
\includegraphics[width=2.5in]{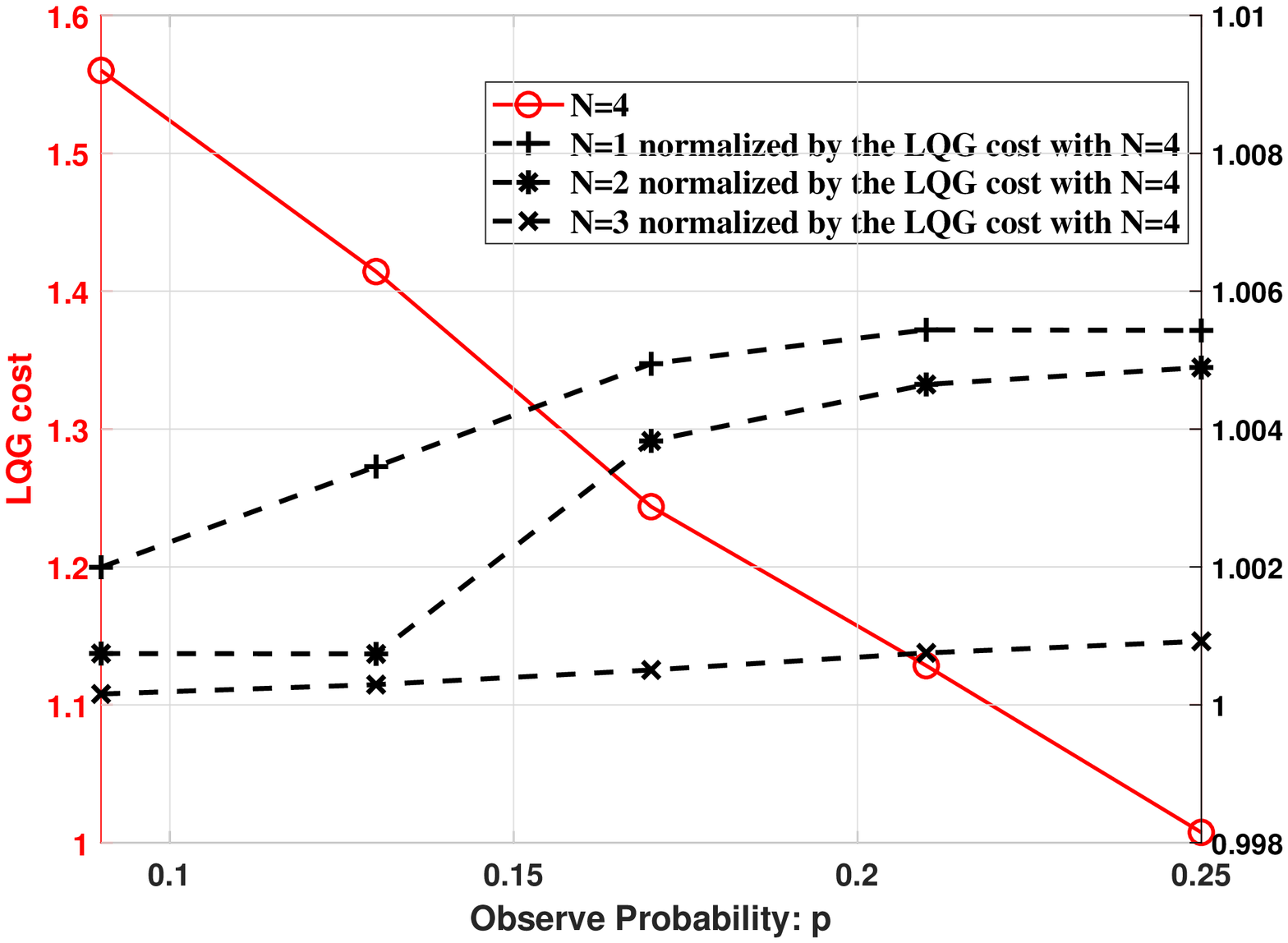}
\end{minipage}%
}%
\centering
\caption{Fig.\ref{fg:sub4} shows the difference between our proposed function and the function in \cite{singh2019optimal}. Fig.\ref{fg:sub1} shows the performance comparison between different policies varying the observation noise variance ${\sigma}_{o,1}^2$. Fig.\ref{fg:sub2} shows the convergence for window size $N$. }
\vspace{-6mm}
\end{figure*}

\vspace{-1mm}
We consider WNCSs with delayed noisy observations updated at the controller in each slot and compare 
the evaluation accuracy of the estimation error with different process noise variance $\{0.01,0.02,0.05\}$ and different delays of observations.  Fig.\ref{fg:sub4} shows the squared estimation error evaluation normalized by the result from Theorem 1. The parameters are set that $\sigma^2_{o}(t) = 0.05$, $a=1.3$, $b=1$, $k=0.5$, $T=10000$. It can be seen that the squared estimation error calculated by our function and Monte-Carlo simulation results match perfectly well and that calculated using the method proposed in \cite{singh2019optimal} is an underestimated version (compared with simulation results), and the deviation increases with larger noise variance and delay because $C_e(t,\boldsymbol{\tau}_t)$ increases rapidly with $\tau(t)$ and $a$.

\vspace{-2mm}
\subsection{Performance of Scheduling}
\vspace{-1mm}
$M \!=\! 4$ sensors that share the ideal channel with one packet per time slot bandwidth limit are considered. We set that $a=1.3$, $b=1$, $Q=1$, $R=1$, $\sigma^2_p=0.1$, $t_d=1$. The control strategy can be calculated as $L=-0.8879$ by using \eqref{control1}. 
We assume sensor $m=1$ keeps observing the system state with probability $p_1 = 1$ and all the other sensors share the same observation probability $p$ for simplicity.
The total simulation time slot is set as $T = 10000$ and the initial state is set as $x(0)=1$.
The following comparison policies are used:
\vspace{-1mm}
\begin{itemize}
\item[1)] \emph{Average age-minimal policy} (age-minimal policy)\cite{sun2019age}: The scheduler chooses the sensor with the highest precision among the sensors with the smallest age to transmit.

\item[2)] \emph{Average observation noise variance-minimal policy} (variance-minimal policy): The scheduler chooses the sensor with the highest precision among all available sensors.

\item[3)] \emph{Random policy}: The scheduler chooses randomly.
\end{itemize}

{To achieve better control performance, all policies will not transmit observations that have already been transmitted.}

Fig.\ref{fg:sub1} shows the LQG cost of different policies with respect to the observation noise's variance ${\sigma}^2_{o,1}$. The noise variance of the other three sensors are set as $0.02$, $0.05$, $0.2$ and the observation probability is set as $p=0.4$. 

It can be seen that the performance deteriorates with the increased noise variance, and our proposed policy has great performance advantages compared with the other policies. 
When the observation noise's variance is small, the variance-minimal policy is better than the age-minimal policy. However, when the observation noise's variance exceeds a certain value, the age-minimal policy is better. It is because the variance-minimal policy mainly considers the effect of high precision observations, while age-minimal policy focuses on the latency and chooses more low precision observations. Our proposed algorithm has taken into account the effect of both. 
\vspace{-2mm}
\subsection{Impact of Window Size}
\vspace{-1mm}
Fig.\ref{fg:sub2} shows the LQG cost of our proposed sliding window policy for different window sizes with respect to the observation probability $p$. The observation noise variances are set as $0.8$, $0.02$, $0.05$, $0.2$. The red solid line represents the LQG cost with window size $N=4$, and the black double dash lines represent the performance ratios of window sizes $N=1,2,3$ to the window size $N=4$. {As the observation probability increases}, the performance gradually improves due to the increase of {available observations}. In addition, performance improves as the window size $N$ increases. {Yet, the improvement is not significant at low observation probability due to limited selections of observations. and gradually decreases with the window size. A} small window size can already provide good control performance with low complexity. 
\vspace{-2mm}					
\section{conclusion}
\vspace{-1mm}
We have obtained a direct evaluation function of the estimation error for WNCSs with noisy and delayed observations. A near optimal scheduling policy is proposed to solve the transmission scheduling problem with low complexity and high LQG performance. Simulation results indicated that the proposed policy can significantly outperform the existing policies. {The application of the proposed evaluation method to complicated communication systems, such as packet dropouts and power constraints, will be left for future work. Besides, the generalization to vector-form systems is also of future interest.}
\bibliographystyle{ieeetr}
\vspace{-2mm}
\bibliography{cite}
\end{document}